\input harvmac
\input epsf

\def\ap{\alpha'}
\def\half{{1\over 2}}

\def\sh{\hat{\sigma}}

\def\({\left(}
\def\){\right)}

\Title{}{\vbox{\centerline{Noncommutative KKLMMT Model}}}

\centerline{Qing-Guo Huang}

\bigskip

\centerline{\it School of physics, Korea Institute for Advanced
Study, } \centerline{\it 207-43, Cheongryangri-Dong, Dongdaemun-Gu,
Seoul 130-722, Korea}
\medskip
\centerline{\it Institute of Theoretical Physics, Academia Sinica}
\centerline{\it P. O. Box 2735, Beijing 100080}
\medskip
\centerline{\it The interdisciplinary center of theoretical Studies,
Academia Sinica} \centerline{\it P. O. Box 2735, Beijing 100080}
\medskip
\centerline{\tt huangqg@kias.re.kr}

\bigskip

In the noncommutative space-time the fine tuning in KKLMMT model
can be significantly released and a nice running of spectral index
fitting the WMAP three year data can be achieved. The fitting
results show that the noncommutative mass scale is roughly
$5\times 10^{14}$ Gev. The string mass scale is higher than the
noncommutative scale unless the string coupling is smaller than
$10^{-10}$.

\Date{May, 2006}


\nref\infl{A.H. Guth, Phys.Rev.D 23(1981)347; A.D. Linde,
Phys.Lett.B 108(1982)389; A. Albrecht and P.J. Steinhardt,
Phys.Rev.Lett. 48(1982)1220. }

\nref\wmapts{D.N. Spergel et. al, astro-ph/0603449.}

\nref\al{L. Alabidi and D.H. Lyth, astro-ph/0603539. }

\nref\pe{H. Peiris., R. Easther, astro-ph/0603587; R. Easther, H.
Peiris, astro-ph/0604214. }

\nref\mr{J. Martin, C. Ringeval, astro-ph/0605367. }

\nref\dt{G.R. Dvali and S.H. Tye, Phys.Lett.B 450 (1999) 72,
hep-ph/9812483.}

\nref\brinf{S. Alexander, Phys.Rev.D 65 (2002) 023507,
hep-th/0105032; G. Dvali, Q. Shafi and S. Solganik, hep-th/0105203;
C. Burgess, M. Majumdar, D. Nolte, F. Quevedo, G. Rajesh and R.J.
Zhang, JHEP 07 (2001) 047, hep-th/0105204; G. Shiu and S.H. Tye,
Phys.Lett.B 516 (2001) 421, hep-th/0106274; D. Choudhury, D.
Ghoshal, D.P. Jatkar and S. Panda, hep-th/0305104. }

\nref\fq{F. Quevedo, hep-th/0210292. }

\nref\kklt{S. Kachru, R. Kallosh, A. Linde and S. Trivedi,
Phys.Rev.D 68(2003)046005, hep-th/0301240. }

\nref\kklmmt{S. Kachru, R. Kallosh, A. Linde, J. Maldacena, L.
McAllister and S. Trivedi, JCAP 0310(2003)013, hep-th/0308055. }

\nref\ft{H. Firouzjahi and S.H. Tye, hep-th/0501099. }

\nref\hk{Qing-Guo Huang and Ke Ke, Phys.Lett.B 633(2006)447,
hep-th/0504137. }

\nref\sh{S. Shandera, S.H. Tye, hep-th/0601099.}

\nref\bcsq{C.P. Burgess, J.M. Cline, H. Stoica, and F. Quevedo,
JHEP 0409(2004)033, hep-th/0403119; J.M. Cline, H. Stoica,
Phys.Rev.D 72(2005)126004, hep-th/0508029. }

\nref\hs{Qing-Guo Huang, Miao Li and Jian-Huang She,
hep-th/0604186.}

\nref\sst{N. Seiberg, L. Susskind and N. Toumbas, JHEP
0006(2000)021, hep-th/0005040. }

\nref\gmms{R. Gopakumar, J. Maldacena, S. Minwalla, A. Strominger,
JHEP 0006(2000)036, hep-th/0005048.}

\nref\ncst{T. Yoneya, in " Wandering in the Fields ", eds. K.
Kawarabayashi, A. Ukawa ( World Scientific, 1987), p. 419; M. Li and
T. Yoneya, Phys. Rev. Lett. 78 (1977) 1219; T. Yoneya, Prog. Theor.
Phys. 103, 1081 (2000), hep-th/0004074; J. Polchinski, " String
Theory " volume 2; J. She, hep-th/0509067, hep-th/0512299.}

\nref\bh{R. Brandenberger, P.M. Ho, Phys.Rev.D 66(2002) 023517,
hep-th/0203119. }

\nref\hla{Q.G. Huang and M. Li, JHEP 0306(2003)014,
hep-th/0304203. }

\nref\hlb{Q.G. Huang and M. Li, JCAP 0311(2003)001,
astro-ph/0308458. }

\nref\hlc{Q.G. Huang and M. Li,  Nucl.Phys.B 713(2005)219,
astro-ph/0311378. }

\nref\hld{Q.G. Huang and M. Li, astro-ph/0603782.}

\nref\zw{X. Zhang and F. Wu, astro-ph/0604195. }

\nref\pncf{S. Tsujikawa, R. Maartens and R. Brandenberger,
astro-ph/0308169; E. Grezia, G. Esposito, A. Funel, G. Mangano, G.
Miele, gr-qc/0305050; M. Fukuma, Y. Kono, A. Miwa, hep-th/0307029;
D. Liu, X. Li, astro-ph/0402063; H. Kim, G. Lee, Y. Myung,
hep-th/0402018; H. Kim, G. Lee, H. Lee, Y. Myung, hep-th/0402198; R.
Cai, hep-th/0403134; G. Calcagni, hep-th/0406006; S. Alavi, F.
Nasseri, astro-ph/0406477; Y. Myung, hep-th/0407066; G. Barbosa, N.
Pinto-Neto, hep-th/0407111; G. Calcagni, S. Tsujikawa,
astro-ph/0407543; G. Barbosa, hep-th/0408071; D. Liu, X. Li,
hep-th/0409075; K. Bamba, J. Yokoyama, hep-ph/0409237; K. Bamba, J.
Yokoyama, hep-ph/0502244; G. Calcagni, hep-ph/0503044; C-S. Chu, B.
R. Greene and G. Shiu, hep-th/0011241; F. Lizzi, G. Mangano, G.
Miele and M. Peloso, JHEP 0206 (2002) 049, hep-th/0203099. }

\nref\prunt{K. Izawa, hep-ph/0305286; S. Cremonini, hep-th/0305244;
E. Keski-Vakkuri, M. Sloth, hep-th/0306070; M. Yamaguchi, J.
Yokoyama, hep-ph/0307373; J. Martin, C. Ringeval, astro-ph/0310382;
K. Ke, hep-th/0312013; S. Tsujikawa, A. Liddle, astro-ph/0312162; C.
Chen, B. Feng, X. Wang, Z. Yang, astro-ph/0404419; L. Sriramkumar,
T. Padmanabhan, gr-qc/0408034; N. Kogo, M. Sasaki, J. Yokoyama,
astro-ph/0409052; G. Calcagni, A. Liddle, E. Ramirez,
astro-ph/0506558; H. Kim, J. Yee, C. Rim, gr-qc/0506122. }

\nref\pruns{A. Ashoorioon, J. Hovdebo, R. Mann, gr-qc/0504135; G.
Ballesteros, J. Casas, J. Espinosa, hep-ph/0601134; J. Cline, L.
Hoi, astro-ph/0603403. }

\nref\cms{L. Pogosian, H. Tye, I. Wasserman, M. Wyman, Phys.Rev.D
68(2003)023506, hep-th/0304188; L. Pogosian, M. Wyman, I. Wasserman,
astro-ph/0403268. }

\nref\ssm{U. Seljak, A. Slosar, P. McDonald, astro-ph/0604335. }

Inflation \infl\ economically solves the horizon problem, flatness
problem and so on in the hot big bang model. It also predicts a
nearly scale-invariant spectra of primordial scalar and tensor
perturbations. A wide range of astronomical data sets are
consistent with the predictions of the $\Lambda$CDM model \wmapts.
The results of WMAP three data are given in \wmapts. For
$\Lambda$CDM model, WMAP three year data only shows that the index
of the power spectrum satisfies
\eqn\wmapn{n_s=0.951^{+0.015}_{-0.019}.} A red power spectrum is
favored at least at the level of 2 standard deviations. If there
is running of the spectral index, the constraints on the spectral
index and its running are \eqn\arn{n_s=1.21^{+0.13}_{-0.16}, \quad
\alpha_s={dn_s \over d \ln k}=-0.102^{+0.050}_{-0.043}, } and the
tensor-scalar ratio satisfies \eqn\br{r\leq 1.5 \quad (95\%
\hbox{CL}).} Even though allowing for a running spectral index
slightly improves the fit to the WMAP data, the improvement in the
fit can not provide strong evidence to require the running. Many
inflation models have been proposed in the last decade. The
precise observational data has been used to rule out some of them,
see for example \refs{\wmapts-\mr,\hs}. But many inflation models
still survive. Here we need to keep in mind that how to construct
a realistic inflation model in a fundamental theory is still an
open question.

Brane inflation model \refs{\dt-\fq}, which is very appealing in
having such a UV completion, is proposed in string theory after the
introduction of the concept of D-branes therein. However there is an
$\eta$ problem in brane inflation model \fq\ which says the distance
between the brane and anti-brane should be larger than the size of
the extra dimensional space, since the branes are too heavy. A more
realistic model is the so called KKLMMT model \refs{\kklmmt,\ft},
where warping effects are employed to make the brane lighter and
thus solve the $\eta$ problem. However the authors in \hs\ pointed
out that a stringent fine tuning is still needed in order that
KKLMMT model can fit the WMAP three year data.

In this short note, we investigate KKLMMT model in the
noncommutative space-time. We find that the noncommutative effects
can accommodate the WMAP results with running of the spectral index
and release the fine tuning in KKLMMT model.

Noncommutative geometry can naturally emerge in string theory. In
\sst\ the commutator between space and time coordinates is not
zero if there is an electric field on the brane, which says
\eqn\nst{[t,x]=i\theta=iM_{nc}^{-2},} where $\theta={1 \over
E_{cr}}{{\tilde E} \over 1-{\tilde E}^2}$, ${\tilde E}=E/E_{cr}$
and $E_{cr}=1/(2\pi \ap)$ is the critical electric field. Beyond
that value strings can materialize out of the vacuum, stretch to
infinity and destabilize the vacuum \refs{\sst,\gmms}. This
noncommutativity leads to an uncertainty relation between time and
space, which was advocated as a generic property of string theory
even when no electric field is present \ncst.

In a quantum theory, time coordinate labels the evolution of the
system. We don't know how the time could fail to commute. Here we
adopt the strategy proposed in \bh\ to explore how the space-time
noncommutative effects quantitatively modify the evolution of the
quantum fluctuations during the period of inflation. There are
many discussions about the parameter space of the KKLMMT model,
see for example \refs{\ft,\sh,\bcsq}. In this short note, we
parameterize KKLMMT model with potential \eqn\gkp{V=\half \beta
H^2 \phi^2+2T_3h^4\(1-{M^4 \over \phi^4} \), } here
\eqn\dm{M^4={27\over 32\pi^2}T_3h^4.} The inflation is governed by
the effective D3-brane tension on the brane \eqn\td{{\tilde
{T_3}}=T_3h^4={M_{obs}^4 \over (2\pi)^3g_s}, } where $h$ is the
warped factor in the throat and $M_{obs}=M_sh$ is the effective
string scale on the brane \hk. The warped factor makes D3-brane
lighter. The $\beta$ term comes from the Kahler potential, D-term
and also interactions in the superpotential, and $H$ is the Hubble
constant. Generally $\beta$ is of order unity \kklmmt, but to
achieve slow roll, it has to be fine-tuned to be much less than
one \refs{\ft, \hk,\hs} and it seems quite unnaturally. On the
other hand, in \refs{\hla-\hlc}, the authors find the space-time
noncommutative effects can accommodate a large enough running. See
\refs{\hld,\zw} for the recent progresses. This model was later
extensively studied in \pncf. Other models with a large running
are discussed in \refs{\prunt,\pruns}.

The spacetime noncommutative effects are encoded in a new product
among functions, namely the star product, replacing the usual
algebra product. The evolution of the background is homogeneous
and the standard cosmological equations of the inflation will not
change. The value of $\phi$, namely, $\phi_N$ at the number of
e-folds equals $N$ before the end of inflation is
\eqn\pn{\phi_{N}^6=24 N M_p^2 M^4 m(\beta), } where
\eqn\mb{m(\beta)={(1+2 \beta)e^{2 \beta N} -(1+\beta/3) \over 2
\beta (N+5/6) (1+\beta/3)}. } Now the slow roll parameter can be
expressed as \eqn\srp{\eqalign{\epsilon_v&={1 \over 18}
\left({\phi_{N} \over M_p} \right)^2 \left(\beta+{1 \over 2 N
m(\beta)} \right)^2, \cr \eta_v&={\beta \over 3} -{5 \over 6}{1
\over N m(\beta)}, \cr \xi_v&={5 \over 3Nm(\beta)}\(\beta+{1 \over
2Nm(\beta)} \). }} The amplitude of the primordial scalar power
spectrum in noncommutative space-time takes the form, (see \hlc\
in detail) \eqn\asd{\Delta_{\cal R}^2\simeq {V/M_p^4 \over
24\pi^2\epsilon_v}(1+\mu)^{-4}=\(2^5 \over 3\pi^4\)^{1/3}
\({T_3h^4\over M_p^4}\)^{2 \over 3} N^{5\over 3}f^{-{4 \over
3}}(\beta)(1+\mu)^{-4}, } where \eqn\fb{f(\beta) =
m^{-{5/4}}(\beta)(1+2\beta N m(\beta))^{3\over 2}, } $\mu = {H^2
k^2 / (a^2 M_{nc}^4)}$ is the noncommutative parameter, $H$ and
$V$ take the values when the fluctuation mode $k$ crosses the
Hubble radius, $k$ is the comoving Fourier mode and $M_{nc}$ is
the noncommutative mass scale. The factor $(1+\mu)^{-4}$ in eq
\asd\ comes from the space-time noncommutative effects.
Substituting eq. \dm\ and \pn\ into \asd, we have
\eqn\pmp{{\phi_{N} \over M_p}=\left({27 \over 8} \right)^{1\over
4} m^{1 \over 6}(\beta) f^{1 \over 3} (\beta) N^{-{1\over 4}}
\(\Delta_{\cal R}^2\)^{1\over 4}(1+\mu). } The normalization of
the primordial scalar power spectrum is $\Delta_{\cal R}^2 \simeq
2\times 10^{-9}$ for $N\sim 50$. Now we have
\eqn\epp{\epsilon_v={1 \over 4\sqrt{6N}}\(\Delta_{\cal
R}^2\)^\half m^{1\over 3}(\beta)f^{2\over 3}(\beta)\(\beta+{1\over
2Nm(\beta)}\)^2(1+\mu)^2.} The spectral index and its running are
\eqn\ind{\eqalign{n_s&=1-6\epsilon_v+2\eta_v+16\epsilon_v \mu, \cr
\alpha_s&=-24\epsilon_v^2+16\epsilon_v\eta_v-2\xi_v-32\epsilon_v
\eta_v \mu,}} with the tensor-scalar ratio
\eqn\tsr{r=16\epsilon_v.}

The noncommutative effects encode in the parameter $\mu$ and these
effects are negligible if $\epsilon_v$ is too small. For
$\beta<0.1$, the tensor perturbations can be negligible \hs\ and the
noncommutative effects can not improve the fitting for the large
running. When $\beta$ becomes large, $\epsilon_v$ becomes also
large. For instance, the spectral index and its running for
$\beta=0.25$ are showed in fig. 1.
\bigskip
{\vbox{{
        \nobreak
    \centerline{\epsfxsize=8cm \epsfbox{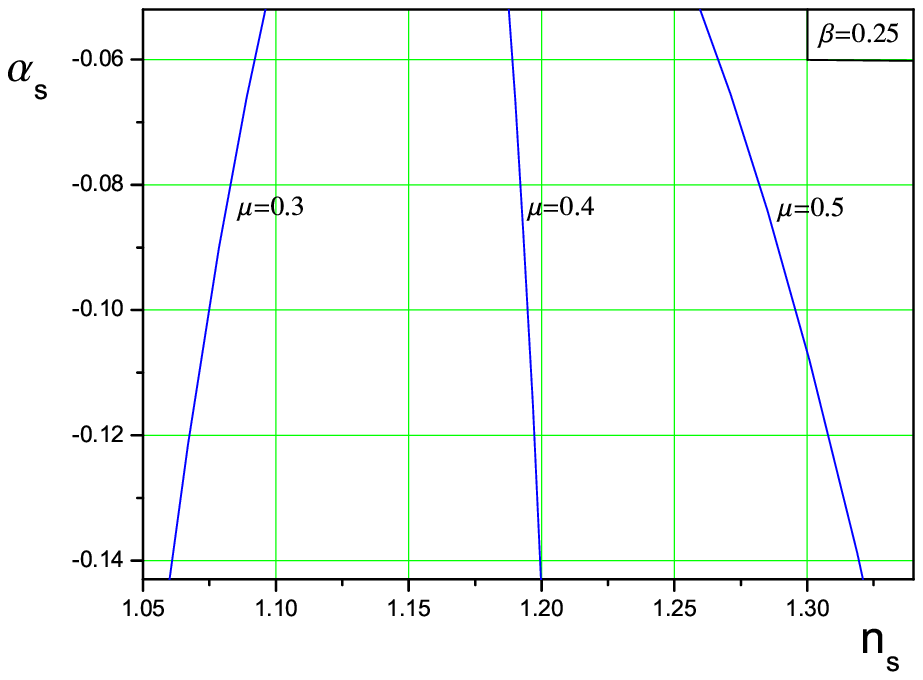}
    }
        \nobreak\bigskip
    {\raggedright\it \vbox{
{\bf Figure 1.} {\it The range for the spectral index and its
running is allowed by WMAP three year data at the level of
$1\sigma$.}
 }}}}
    \bigskip}
We also take $\beta=0.25$ and $\beta=0.3$ to fit three year results
of WMAP respectively. The fitting results for eq. \arn\ are
\eqn\nnn{\eqalign{N&=47.7^{+1.6}_{-3.7}, \quad
\mu=0.414^{+0.241}_{-0.157}, \quad r=1.12^{+0.31}_{-0.50},\quad
\hbox{for} \quad \beta=0.25, \cr N&=38.4^{+1.4}_{-2.8}, \quad
\mu=0.384^{+0.204}_{-0.148}, \quad r=1.17^{+0.33}_{-0.51},\quad
\hbox{for} \quad \beta=0.3,}} at the level of one stand deviation.
The allowed range for $\mu$ and the number of e-folds $N$ is showed
in fig. 2 where the range of $\beta$ we scan is $0.22\leq \beta \leq
0.31$.
\bigskip
{\vbox{{
        \nobreak
    \centerline{\epsfxsize=8cm \epsfbox{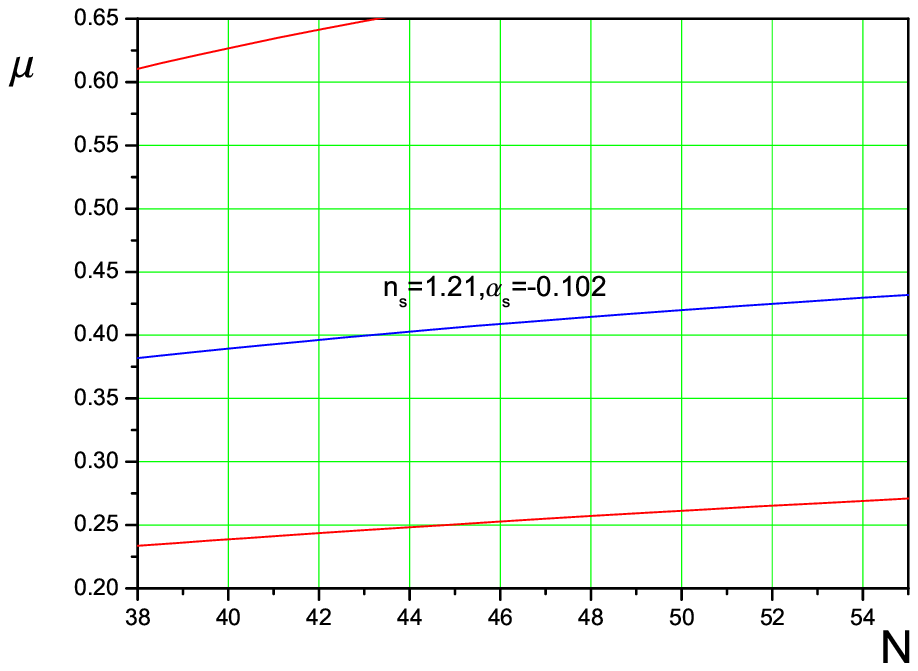}
    }
        \nobreak\bigskip
    {\raggedright\it \vbox{
{\bf Figure 2.} {\it Here the range of $\beta$ is $0.22\leq \beta
\leq 0.31$. The range between the two red lines is allowed by the
WMAP three year data at the level of $1\sigma$. The blue line
corresponds to $n_s=1.21$ and $\alpha_s=-0.102$.}
 }}}}
    \bigskip}

The space-time noncommutative effects can improve KKLMMT model to
nicely accommodate the spectral index and its running. Requiring
$N\geq 47$ yields $\beta\leq 0.25$.

Before the end of this note, we also want to work out the
noncommutative mass scale. For $N=50$, we read out that $\beta=0.24$
and $\mu=0.42$ corresponding to the blue line in fig. 2. Using eq.
\asd, we can decide the value of the effective D3-brane tension as
\eqn\bttm{{\tilde {T_3}}\sim 7\times 10^{-8} M_p^4, \quad \hbox{or}
\quad {M_{obs}^4\over g_s} \sim 2\times 10^{-5}M_p^4.} In order that
the noncommutative effects become significant, the noncommutative
mass scale is roughly the same as the Hubble constant during the
period of inflation, namely $\mu$ is not quite smaller than one or
$M_{nc} \sim H \sim \sqrt{2{\tilde{T_3}}\over 3M_p^2}\sim 2\times
10^{-4}M_p \simeq 5\times 10^{14}$ Gev. If $g_s\simeq 8\times
10^{-11}$, $M_{nc}=M_{obs}$. We don't expect the string coupling is
so small and thus the string mass scale is higher than the
noncommutative mass scale.

In brane inflation, cosmic strings are possibly generated after
the end of the inflation. Fitting cosmological constant plus cold
dark matter plus strings to the CMB power spectrum provides an
upper limit on the string tension with $GT \leq 10^{-6}$ in \cms\
and recent constraint $GT \leq 2.3\times 10^{-7}$ in \ssm\ (95$\%$
CL), where $T$ is the cosmic string tension. For D-string,
$GT_D=({1\over 32\pi g_s}{{\tilde{T_3}}\over M_p^4})^{1/2}\sim
3\times 10^{-5}/\sqrt{g_s}$. If cosmic D-sting was produced, $g_s
\geq 10^4$. For fundamental string, $GT_F={1\over
8\pi}\({M_{obs}\over M_p}\)^2\simeq 2\times 10^{-4}\sqrt{g_s}$. If
cosmic fundamental string was generated, $g_s<2\times 10^{-6}$.

To summarize, a nice running of the spectral index is obtained and
the fine tuning for the parameter $\beta$ is significantly
released in the noncommutative KKLMMT model. The noncommutative
mass scale is roughly $5\times 10^{14}$ Gev which can be different
from the string scale. If cosmic strings are produced after
inflation, the constraint on the string coupling becomes
stringent. We expect the future cosmological observations can
provide stronger evidence to support a large amplitude of the
tensor perturbations and a running of the spectral index for the
primordial scalar power spectrum.

\bigskip

Acknowledgments.

We thank M. Li and J.H. She for useful discussions. This work in
part was supported by a grant from NSFC, a grant from China
Postdoctoral Science Foundation and and a grant from K. C. Wang
Postdoctoral Foundation.

\listrefs
\end